\documentclass[aps,prd,floatfixy,superscriptaddress,showpacs,showkeys]{revtex4}[18pt]
\usepackage{amssymb}
\usepackage{amsmath}
\usepackage{amsfonts}
\usepackage{epsfig}
\usepackage{graphicx}
\usepackage{graphics}
\usepackage{subfigure}
\usepackage{color}
\usepackage{verbatim}
\def\baselinestretch{1.2}

\newcommand{\eq}{\begin{eqnarray}}
\newcommand{\en}{\end{eqnarray}}

\usepackage[normalem]{ulem}

\renewcommand\sout{\bgroup \color{red} \ULdepth=-.5ex \ULset}

\definecolor{orange}{RGB}{255,127,0}
\definecolor{pink}{RGB}{255,192,203}
\definecolor{brown}{RGB}{139,35,35}
\definecolor{Magenta}{RGB}{255,0,255}
\def\ds{d^*(2380)}
\def\eq{\begin{eqnarray}}
\def\en{\end{eqnarray}}
\def\ds{d^*}
\def\d12{D_{12}}

\begin{document}

\title{\huge\bf  Photo-absorption on deuteron \\
contributed by $d^*(2380)$ resonance}
\author{Yubing Dong}
\affiliation{Institute of High Energy Physics, Chinese Academy of
Sciences, Beijing 100049, China}
\affiliation{Theoretical Physics
Center for Science Facilities (TPCSF), CAS, Beijing 100049, China}
\affiliation{School of Physical Sciences, University of Chinese
Academy of Sciences, Beijing 101408, China}
\author{Pengnian Shen}
\affiliation{College of Physics and Technology, Guangxi Normal
University, Guilin  541004, China}
\affiliation{Institute of High
Energy Physics, Chinese Academy of Sciences, Beijing 100049, China}
\affiliation{Theoretical Physics Center for Science Facilities
(TPCSF), CAS, Beijing 100049, China}
\author{Zongye Zhang}
\affiliation{Institute of High Energy Physics, Chinese Academy of
Sciences, Beijing 100049, China} \affiliation{Theoretical Physics
Center for Science Facilities (TPCSF), CAS, Beijing 100049, China}
\affiliation{School of Physical Sciences, University of Chinese
Academy of Sciences, Beijing 101408, China}
\date{\today}
\begin{abstract}
In order to understand the possible physical nature of the newly observed resonance $d^*(2380)$, we calculate
the real photo-absorption cross section on deuteron contributed by the resonance $\ds$ by considering the
electromagnetic transition amplitude of $\gamma +d\rightarrow d^*(2380)$. In our interpretation, the $d^*(2380)$ is
regarded as a compact six-quark system with mainly two components of $\Delta\Delta$ and hidden-color clusters
$C_8C_8$. We find that only the next-to-leading terms contribute the $\gamma +d\rightarrow d^*(2380)$ and the obtained
photo-absorption cross section is quite small which is in the order of 10 $nb$. Compared with data measured at ELPH
and Mainz recently, it is almost about 20 times smaller.
\end{abstract}
\maketitle
\section{Introduction}
\noindent\par

Since dibaryon states were proposed more than half century ago, their existence has become one of the most interest
issues of hadronic physics. Among various dibaryon states, $H$ particle and $\ds$ were involved most. In particular,
the $d^*$ state has been explicitly studied by different approaches from the hadronic degrees of freedom to the quark
degrees of freedom, and the obtained binding energy was ranged from a few MeV to several hundred MeV. Searching for
such an interesting state has also been considered as one of the aims in several experimental projects. However, no
convincing results were released until 2009. After that, a series of experimental studies for $\ds$ was carried out
in the analysis of ABC effect by CELSIUS/WASA and WASA@COSY Collaborations
~\cite{CELSIUS-WASA,Adlarson:2011bh,Adlarson:2012fe,Adlarson:2014pxj}.
Various double-pion and single-pion decays, including invariant mass spectra, Dalitz plots, Argon plots, in the $pn$
and $pA$ reactions, the analyzing power of the neutron-proton scattering and etc., have been  measured and analyzed.
It was found that the experimental results cannot be simply understood by the contribution either from the intermediate
Roper excitation or from the t-channel $\Delta\Delta$ effect, except introducing an intermediate new resonance. Then,
the discovery of a new resonance, with a mass (width) about $2370\sim 2380~\rm{MeV}$ ($70\sim 80~\rm{MeV}$) and the
quantum numbers of $I(J^P)=0(3^+)$, was announced
~\cite{CELSIUS-WASA, Adlarson:2011bh, Adlarson:2012fe, Adlarson:2014pxj}. It is believed that such a state is just
the $d^*$ state which has been hunted for several decades due to its baryon number being 2. In general, it can be
explained by either "an exotic compact particle" or "a hadronic molecule state" (see the review article of
Ref.~\cite{Clement:2016vnl}). \\

One may reasonably expect that the threshold (or cusp) effect may not be so significant in the $\ds$ case as in the
XYZ particles due to the fact that the observed mass of $\ds$ is about $80~\rm{MeV}$ below the $\Delta\Delta$
threshold and about $70~\rm{MeV}$ above the $\Delta\pi N$ threshold~\cite{Chen:2016qju, Guo:2017jvc,Dong:2017gaw}.
If $d^*$ does exist, it contains at least 6 light quarks, and it is also much different from the XYZ particles which
contain heavy flavor.\\

Up to now, many theoretical models for the structure of $d^*$ have been developed or proposed.  There are mainly two
structural schemes which attract considerable attention of community. One  assumes that the $d^*$ state has a
compact structure, and may be an exotic hexaquark dominated state whose mass is about  $2380-2414~\rm{MeV}$ and width
about $71~\rm{MeV}$, respectively ~\cite{Yuan,Brodsky,Huang:2014kja,Huang:2015nja,Dong,Dong1,Dong2,Dong2017}.
Some quark models calculations for the dibaryon $d^*(2380)$ are also referred to~\cite{Ping:2008tp,Huang:2013rla}.
The other one, in order to explain the upper limit of the single-pion decay width of $\ds$~\cite{Clement2017},
proposes that the $d^*$ state is basically a molecular-like hadronic state~\cite{Gal1}, which originates from a
three-body $\Delta N \pi$ resonance assumption, where the pole position of the resonance locates around ($2363\pm
20$)~$+{\it i} (65\pm 17)$ MeV~\cite{Gal:2013dca,Gal:2014zia}, and a $D_{12}\pi$ molecular-like model, where
the mass and width of the resonance are pre-fixed to be $2370~\rm{MeV}$ and $70~\rm{MeV}$, respectively
~\cite{Platonova:2014rza,Platonova:2012am}. Although some of the experimental data, like its mass and double
pion decays, can be explained by using either scheme, the described structures of $\ds$ are quite different.
Therefore, it is necessary to seek some other physical observables which would have distinct values for
the different interpretations so that with the corresponding measured experimental data one would be able to
justify which one is more reasonable.\\

It is known that the electromagnetic form factors are the indispensable physical quantities to show the internal
structure of a complicated system. The electromagnetic form factors of a nucleon, for example, provide the charge
and magnetic distributions inside the nucleon. The accurately measured charge radius of the proton may justify
the structure of the nucleon. Consequently, the electromagnetic form factors of a the higher spin particle are
also a  discriminating quantity for different approaches. In particular, for the $\ds$ state, if there is a
considerably large hidden-color component (HCC) in it, we have found that, although such a component does not
contribute to its hadronic strong decay in the leading-order calculation, it plays a rather important role in the
charge distribution calculation~\cite{Dong,Dong1,Dong2,Dong:2017mio}, and the obtained charge distribution with a
compact 6-quark scenario is quite different from that having a $D_{12}\pi$ (or $\Delta\pi N$) structure
~\cite{Dong:2017mio,Dong:2018emq}. Other physical quantities, like the $\ds$ production in the $e^+e^-$
annihilation, and constituent quark counting rule in the high energies may also provide some other information
for its structure~\cite{Lu:2018gtk,Dong:2019stt}. \\

Another physical quantity to explore the structures of the nucleon excitations, is the electromagnetic transition
amplitudes in the $\gamma$-nucleon ($\gamma$-N) interaction, such as the $\gamma N\to N^*$ process. There are many
calculations for the electromagnetic transition amplitudes of the nucleon excitations in the $\gamma N$ process,
for example in $\gamma~N\to \Delta(1232)$, $\gamma N\to S_{11}(1535)$, and $\gamma N\to D_{13}(1520)$ et al.
~\cite{Close:1989aj,Li:1990qu,Giannini:1990pc,Capstick:1992xn,Capstick:1992uc,Pascalutsa:2006up,Santopinto:2012nq,
Warns:1989xr,Warns:1989ie,Bijker:1994yr,Ferraris:1995ui,Dong:2001js}. Those amplitudes can also tell the nature of
those nucleon resonances. However, to extract them, one has to measure the physical process, such as pion
photo-production, or meson photo-production~ \cite{Chiang:2001as,Kamalov:2001qg,Yang:1985yr,Yang:1989si,
Nozawa:1989gy}. It is also possible to get the information of the transition amplitudes from the photo-absorption
cross section on the nucleon target ~\cite{Stoler:1993yk,Stuart:1996zs,Dong:1997pv,Drechsel:2004ki,Aznauryan:2009da}.
There are some experimental data on the cross sections of the photo-absorption for the $^1H$, $^2H$, and $^3He$
~\cite{MacCormick:1996jz, MacCormick:1997ek} targets and also for some nuclei of $Li$, $Be$, $C$, $Al$, $Pb$, $Sn$,
and $U$~\cite{Bianchi:1994ax,Bianchi:1995vb,Arndt:2005wk}, which can tell the nuclear medium effects.\\

Analogy to the study of the transverse helicity amplitudes, like $A_{1/2}$ and $A_{3/2}$ of $\Delta$ resonance in
$\gamma+N\to\Delta$, here, we show a model-dependent calculation for the real photo-absorption cross section on the
deuteron at the $\ds$ energy region. The contribution by the $\ds$ resonance for $\gamma +d\to d^*$, which is directly
associated with the matrix elements of the electromagnetic interaction, is explicitly shown. It should be
stressed that in the single baryon (three-quark system) case, the transition amplitudes are contributed by tree (or
leading) diagrams due to the photon-quark coupling. But, in the $\ds$ case, the amplitudes are obtained from the
contributions of sub-leading diagrams, which is very similar to those in the case of the single-pion decay of
$\ds$~\cite{Dong2017} due to the fact that the deuteron is mainly composed of a proton and a neutron, whereas the
$d^*(2380)$ resonance is {made up of about $31\%$} $\Delta\Delta$ configuration and about $69\%$ $C_8C_8$
hidden-color configuration in our scenario. It should be mentioned that a very recent analysis~\cite{Bashkanov:2019mbz}
shows that the experiment of intense photoinduced reaction on deuteron can provide abundant information for the
structural characteristics, such as the size, the magnetic dipole and quadrupole moments, as well the deformation,
of $\ds$.\\

This paper is organized as follows. In Sect. II, the hypothetic structure and the corresponding wave function
of the $\ds$ resonance in the extended chiral SU(3) constituent quark model is briefly introduced. Sect. III
is devoted to the calculation of the matrix elements of the electromagnetic interaction for the process
$\gamma+d\to \ds$. The relevant photo-absorption cross section on deuteron contributed by $\ds$ is
given in Sect. IV. Finally a short summary is presented in Sect. V.\\

\section{Structure and wave function of $d^*(2380)$ in the extended chiral constituent quark model}
\par\noindent\par

In 1999, a $\Delta\Delta+C_8C_8$ structure of the $\ds$ state with $\big(I(J^P))=(0(3^+)\big)$ (where $I$, $J$, $P$
are isospin, spin, and parity of the system, respectively,) was proposed in Ref.~\cite{Yuan} and its binding
energy and root mean square radius (RMS) were predicted. Recently, a series of sophisticated calculations on
the structure and decay characteristics of $\ds$ had further been performed, and the obtained mass, all the
partial decay widths and total width of $\ds$ are all consistent with the observed data. Then, a picture of a
compact structure, an exotic hexaquark dominated state, was deduced
~\cite{Yuan, Huang:2014kja,Huang:2015nja,Dong,Dong1,Dong2,Dong:2017mio}. In order to make this conclusion more
meaningful, a so-called extended chiral SU(3) constituent quark model (ECCQM) that provides the basic effective
quark-quark interactions caused by the exchanges of the chiral fields, including pseudo-scalar, scalar and vector
mesons, and one gluon, as well as by the quark confinement, was employed in the dynamical calculations and in the
quark degrees of freedom. In terms of this ECCQM,  in which the model parameters are determined by the stability
conditions and the masses of the ground state baryons, the static properties of baryons, the binding energy of
deuteron, the phase shifts of the $N$-$N$ scattering and the cross sections of the N-hyperon (N-Y) interactions
can be well reproduced showing the predictive power of ECCQM~\cite{Yu:1995ag,Zhang:1997ny}.\\

Specifically, the structural calculation was carried out in the well-established Resonating Group Method (RGM),
which has frequently been applied to the studies of nuclear physics and hadronic physics, especially where
the clustering phenomenon does exist~\cite{Tang:1977tw,Oka:1981rj,Faessler:1983yd,Lacombe:2002di,
Faessler111320,Faessler111321,Faessler111322,Oka1984,Shimizu:1989ye,Yamauchi:1992fh}. In our compact structure, the
trial wave function of this six-quark system with two-configuration, $\Delta\Delta$ and $C_8C_8$, can be written as
\begin{eqnarray}\label{Eq:wavfun1}
\Psi_{6q}&=&{\cal A} \Big [  \hat{\phi}^A_\Delta\!\left(
{\vec{\xi}}_1, {\vec{\xi}}_2, \mu_{\Delta}^{A} \right)
\hat{\phi}^B_\Delta\!\left( {\vec{\xi}}_4, {\vec{\xi}}_5,
\mu_{\Delta}^{B}\right) \eta_{\Delta\Delta}\! \left({\vec{r}}\right)\\ \nonumber
&&+\hat{\phi}^A_{\rm C_8}\!\left( {\vec{\xi}}_1, {\vec{\xi}}_2,
\mu_{\rm C_8}^{A}\right) \hat{\phi}^B_{\rm {C_8}}\!\left(
{\vec{\xi}}_4, {\vec{\xi}}_5, \mu_{\rm C_8}^{B}\right) \eta_{\rm
C_8C_8}\! \left({\vec{r}}\right) \Big ]_{S=3,
T=0}^{C=\left(00\right)},
\end{eqnarray}
where $S$, $T$ and $C$ represent the quantum numbers of the spin, isospin and color, ${{\cal A}=
1-\hspace{-0.5cm}{\sum\limits_{i(\in A),~j(\in B)}}\hspace{-0.5cm}P_{ij}^{OSFC}}$ is an
anti-symmetrization operator with $P_{ij}^{OSFC}$ denoting the exchange operator which exchanges the
$i$-th quark belonging to the cluster A and $j$-th quark pertaining to the cluster B in the orbital, spin,
flavor and color spaces, $\hat{\phi}^{A(B)}_{\Delta {\rm(or \, C_8)}}$ depicts the anti-symmetrized internal
wave function of the three quark cluster A(B) for either $\Delta$ or $C_8$ with ${\vec{\xi}}_i$
($i=1,2 ~(4,5)$) being its internal Jacobi coordinates, $\mu_{\Delta {\rm (or \, C_8)}}^{A(\rm B)}$ stands for
an aggregate of the quantum numbers of the spin, isospin and color of the cluster A(B) for either $\Delta$
or $\rm C_8$ with $[{\it S,I,C}]_{\Delta(\rm C_8)} = [ 3/2,3/2,(00)~(\,(3/2,1/2,(11) \,)\,)\,]$ for
the $\Delta(\rm C_8)$ cluster, and $\eta_{\Delta\Delta {\rm (C_8C_8)}}$ is the relative wave function between
the A and B clusters. $\eta_{\Delta\Delta {\rm (C_8C_8)}}$  can be determined by dynamically solving the RGM
equation~\cite{Huang:2014kja,Huang:2015nja}.
\iffalse
\begin{equation}\label{eq:RGM-bound}
\langle \delta\Psi_{6q} | H-E | \Psi_{6q} \rangle = 0
\end{equation}
of the system with ECCQM.  In particular,\fi
The reason for including a $C_8C_8$ hidden-color configuration is that as energy increases, the two $\Delta$ clusters
can get closer, and they may be excited into two-colored clusters, and the two colored cluster can form a color
singlet state. Consequently, such an additional configuration is QCD-allowed in enlarging the Fock space for a
better description of the two-baryon system. In fact, this kind hidden-color configuration has frequently been
employed to study structures of exotic hadrons, such as tetraquarks
~\cite{Brodsky:2014xia,Brodsky:2015wza,Lebed:2018jcr,Brambilla:2014jmp,Maiani:2004vq,Maiani:2014aja,Karliner:2006hf}.\\

However, the resultant wave functions of the two configurations shown in Eq.~(\ref{Eq:wavfun1}) are not orthogonal to
each other. Physically, it means that the obtained $\Delta\Delta$ wave function does not only include the
contributions from the non-hidden-color component, but also contains those from the hidden-color component due to the
antisymmetrization operation. To orthogonalize the wave functions of these two configurations and make the numerical
calculations much simplified and feasible without missing most of the important effect of anti-symmetrization, we
further simplify the RGM wave function of the system by using the channel wave function with the projection
procedure. This method is often used in nuclear physics and hadronization in hadron physics
~\cite{Kusainov91,Glozman93,Stancu97}. Finally, we modify Eq.~(\ref{Eq:wavfun1}) as an effective wave function as
\begin{eqnarray}
\label{eq:wf2} |\ds(S_{\ds}=3,m_{\ds})>&=&
\iffalse
~~~\Big
[|\Delta\Delta>_{M_{\ds}}\chi_{\Delta\Delta}^{S,0}\Big
]_{S_{\ds}=3,M_{\ds}}
+\Big [|\Delta\Delta>_{M_{S}}\chi_{\Delta\Delta}^{D,m_l}
\Big ]_{S_{\ds}=3,M_{\ds}} \nonumber \\
&&+\Big[|C_8C_8>_{M_{\ds}}\chi_{C_8C_8}^{S,0}\Big
]_{S_{\ds}=3,M_{\ds}}+\Big
[|C_8C_8>_{M_S}\chi_{C_8C_8}^{D,m_l}\Big ]_{S_{\ds}=3,m_{\ds}}\nonumber \\
 &=&\fi
\sum_{ch=\Delta\Delta, C_8C_8}~~~\sum_{pw=S,D} \Big
[|ch>_{M_S}\chi_{ch}^{pw,m_l}(\vec{r})\Big
]_{S_{\ds}=3,m_{\ds}}
\end{eqnarray}
with $ch=\Delta\Delta$ and $C_8C_8$ denoting the constituents of the configuration, $m_{\ds}$ representing
the magnetic quantum number of spin $S_{\ds}$, $pw=l=0$ and $2$ depicting the $S$ and $D$ partial waves
($pw$) between the two clusters, respectively, and $m_l$ being its magnetic quantum number. Now, these
four channel wave functions are orthogonal to each other. There are two points should be mentioned: (1)
This treatment is just an approximation. The inaccuracy of such effective wave function is expected to be
about 20\% compared to the wave function obtained in the rigorous RGM dynamical calculation. This is because
that due to the anti-symmetrization procedure, more configurations other than the initially selected $\Delta\Delta$
and $C_8C_8$ that span our model space are generated. Considering the uncertainty in the constituent
quark model caused by non-perturbative QCD (NPQCD), we believe that the contribution from those spare
configurations is not so important. (2) The $D$-wave is ignored in the calculations for the strong decay and
charge distribution, because it is negligibly small. However, it may contribute to the higher multi-pole
form factors, such as $E2$, and $M3$ since those values are closely related to the matrix elements of
the high-rank operators~\cite{Dong:2018emq}. The more detailed information about the wave function of $\ds$
is referred to Ref.~{\cite{Dong:2017mio,Dong:2018ryf}}.\\

\section{Electromagnetic Transition Form Factors of $\gamma+d\to d^*(2380)$}
\noindent\par

When one deals with the transverse transition amplitudes of the nucleon excitations, like $\Delta(1232)$,
$P_{11}(1440)$, $S_{11}(1535)$, and $D_{13}(1520)$, in the process of $\gamma N\to N^*$, the spin projection of
the initial nucleon (spin-1/2 particle) can be antiparallel and parallel to the spin projection of the incoming
photon ($\epsilon_{\gamma}=+1$), and therefore we have two transition amplitudes $A_{1/2}$ and $A_{3/2}$ (or
two helicity amplitudes in the real photon case $Q^2=0$), respectively. Those amplitudes are the matrix elements
of the electromagnetic interaction. When we discuss the electro-production amplitudes, the virtual photon is
considered and the longitudinal transition amplitude $S_{1/2}$ should be included as an additional transition
amplitude, except for the two transverse ones. In the three constituent quark model for baryons, these three
amplitudes are usually calculated in the Breit-frame, and the momentum and energy of the incoming photon are
defined as~\cite{Capstick:1992xn,Capstick:1992uc}
\eq
\label{eq:breitN}
k^2=\frac{(Q^2+M_N^2+M_{X}^2)^2-4M_N^2M_{X}^2}{Q^2+2(M_N^2+M_{X}^2)},~~~~~~
k_0=\sqrt{M_{X}^2+\frac{\vec{k}^2}{4}}-\sqrt{M_N^2+\frac{\vec{k}^2}{4}},
\en
where, $M_N$ and $M_{X}$ are the masses of the nucleon and nucleon excitation, and $Q^2=-(k_0^2-\vec{k}^{~2})$ depicts
the squared momentum transfer.\\

\begin{figure}[htbp]
\begin{center}
\vspace{0.5cm}
\includegraphics[width=6cm,height=3cm] {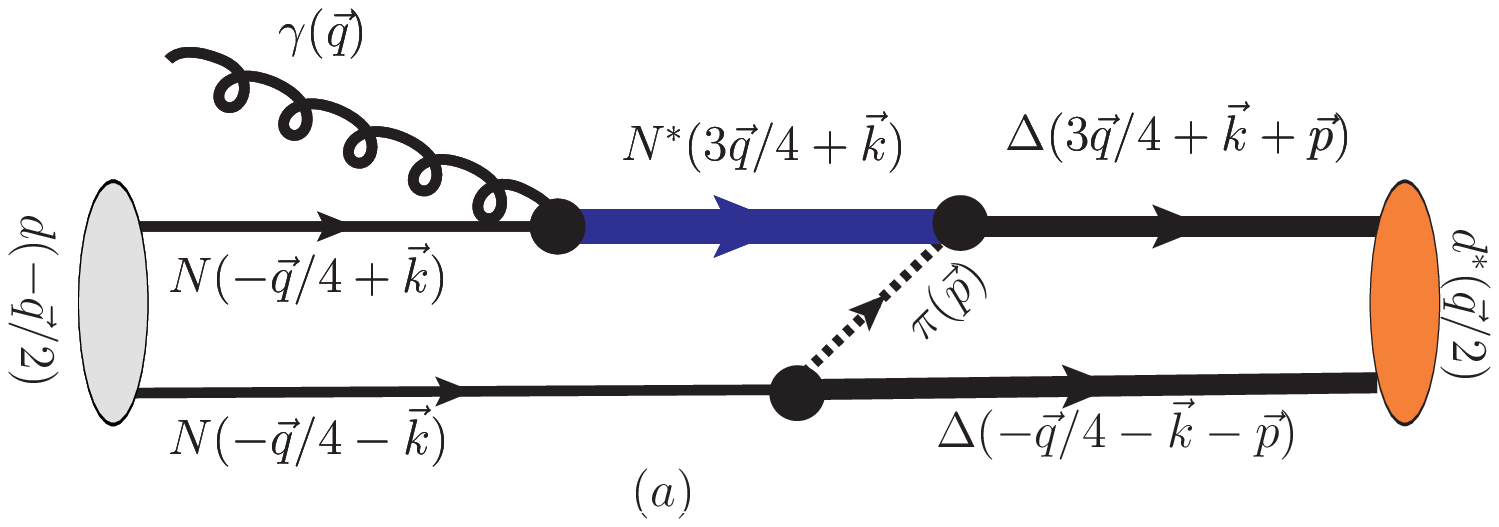}
\vspace{1.cm}
{\hskip 0.5cm}
\includegraphics[width=6cm,height=3cm] {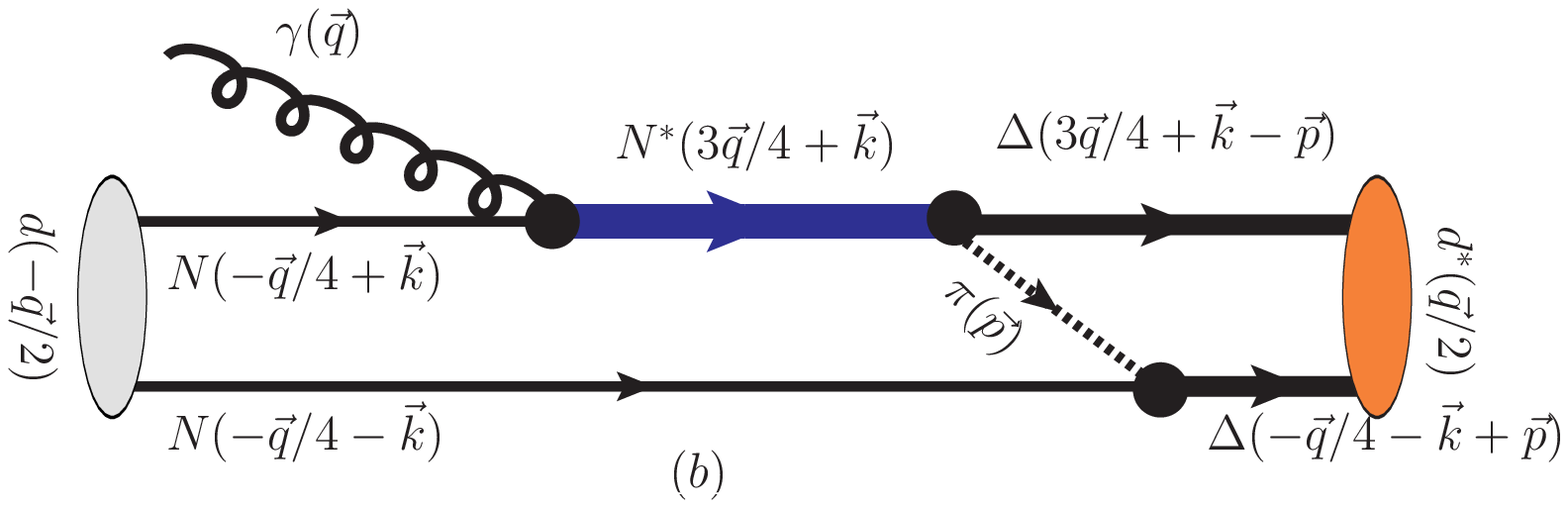}
\vspace{0.5cm}
\includegraphics[width=6cm,height=3cm] {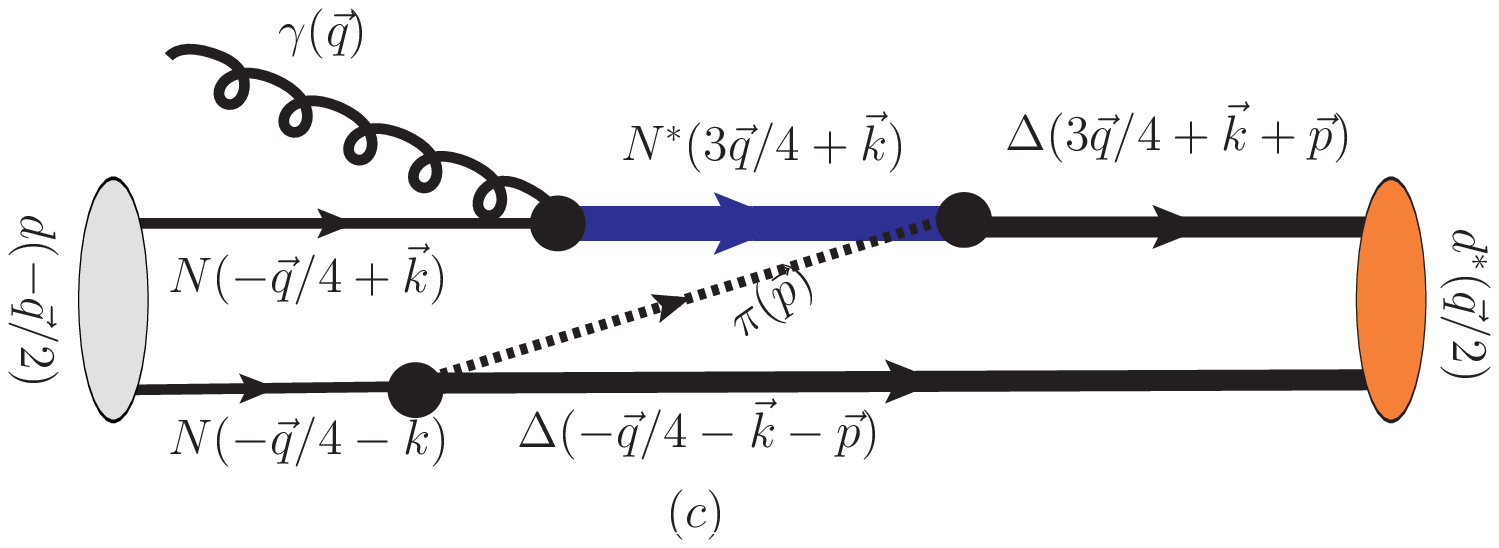}
\vspace{1.cm}
{\hskip 0.5cm}
\includegraphics[width=6cm,height=3cm] {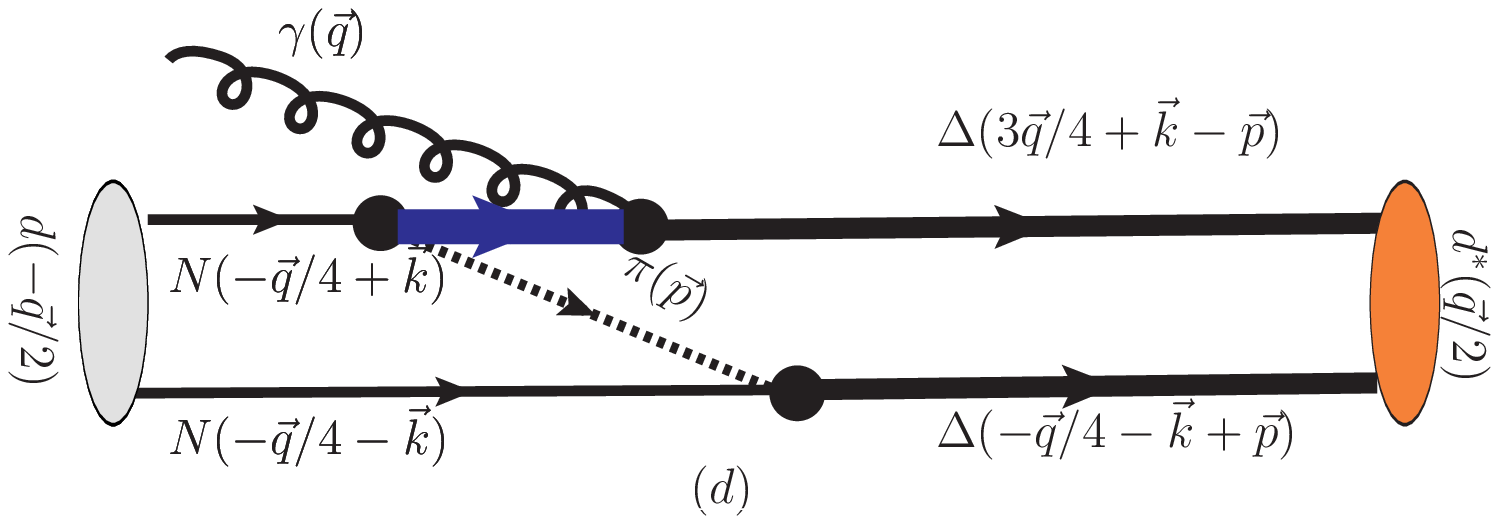}
\vspace{0.5cm}
\includegraphics[width=6cm,height=3cm] {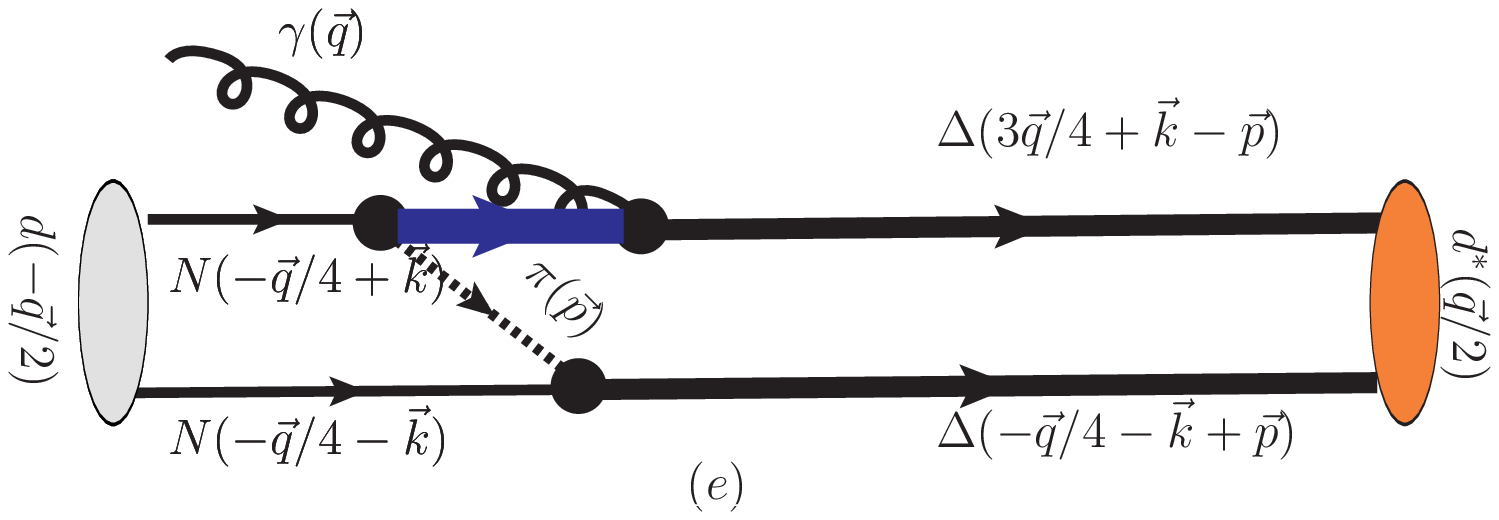}
\vspace {1.0cm}
{\hskip 0.5cm}
\includegraphics[width=6cm,height=3cm] {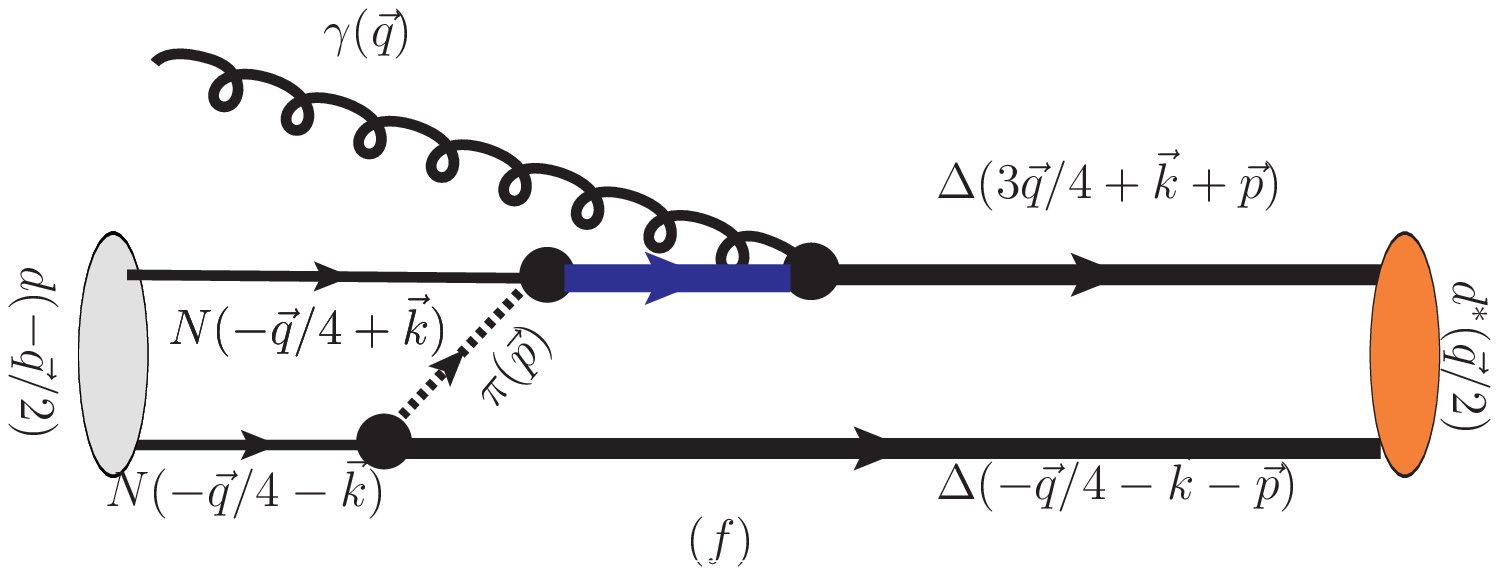}
%{\hskip 1cm}
%{\hskip 1.5cm}
\caption{Feynman diagrams for $\gamma d\to \ds$, where the thick
blue line stands for the intermediate states of the $N$ and
$\Delta$, and the momenta are those in the Breit Frame. Here, only
the diagrams where the photon couples directly to the upper cluster
are explicitly drawn.}
\label{fig:Fig1}
\end{center}
\end{figure}

Analogy to the calculation of transition amplitudes of  $\gamma N \to N^*$, we perform a calculation for the
transition of $\gamma d\to \ds$. Here we employ the Breit frame as well, where the incoming photon has a
four-momentum $k=(k_0,\vec{k})$, then the initial deuteron and final $d^*(2380)$ have three-momentum of
$-\vec{k}/2$ and $\vec{k}/2$, respectively. We also assume that the deuteron can be reasonably regarded as a
weakly bound state of a proton and a neutron, and our $\ds$ is mainly composed of two components, one is a
hidden-color cluster $C_8C_8$ (with a large fraction of about $69\%$) and a $\Delta\Delta$ cluster with a
relatively small fraction of about $31\%$. Therefore, in the calculation of the $\gamma d\to \ds$ transition, or
of the matrix element of the electromagnetic interaction between the initial deuteron and final $\ds$,
photo-deuteron in the leading-order approximation cannot directly reach to the final $\ds$ state, and its
contribution vanishes. Then, we have to consider next to leading-order (NLO) terms, where exist the
intermediate nucleon state and the pion exchange between the two clusters. This feature is similar to our
calculation of the single pion decay partial width of $\ds$~\cite{Dong2017}. The possible diagrams of the NLO
contribution where the photon couples directly to the upper cluster are drown in Fig.~{\ref{fig:Fig1}} and the
momenta are those in the known Breit frame. Of course, the similar diagrams where the photon couples directly
to the lower cluster also contribute, and these contributions are included in the calculation as well.
Figs. 1 (a,b,c) and Figs. 1 (d,e,f) exhibit that the incoming photon acts on the upper cluster before and after
one-pion-exchange occurs, respectively.  \\

In the case of $\gamma +d\to \ds$ where the deuteron is the target, since the initial deuteron is a
massive spin-1 particle with three spin projections, we have three independent transition amplitudes,
two transverse amplitudes with the deuteron polarizations antiparallel and parallel to the incoming real photon,
and one amplitude with the longitudinal polarization of the initial deuteron. Then, the transition amplitudes of the
deuteron contributed by $\ds$ are ${\cal A}(m_d)$ (with $m_d=\pm 1, 0$ being the spin polarizations of the deuteron).
So, in the real photon limit, in terms of our model wave function, we can investigate the helicity amplitudes
$A_{m_d}$, and consequently the total photoabsorption cross section contributed by $\ds$.\\

Of course, the $\gamma +d \to d^*$ process can also be studied in terms of the multipoles. For the $\gamma N\to \Delta$
transition, with $J_i=1/2$ and $J_f=3/2$ being the angular moments of the initial state N and the final state
$\Delta$, we have $M1$ and $E2$ multipole transitions, since the relation of $\mid
J_f-J_i\mid \leq L\leq J_f+J_i$ gives $L=1$ and $2$ with $L$ denoting the order of the multipole transition and
parities of the initial nucleon and final $\Delta$ being both positive (see Refs.
~\cite{Capstick:1994ne,Arenhovel:1990yg,Pascalutsa:2006up} for details).  For the process of $\gamma + d(1^+)\to
\ds(3^+)$ concerned, if we assume $d$ and $d^*$ as point particles, the above
relation gives $L=2$, $3$ and $4$, which link to $E2$, $M3$ and $E4$ multipole transitions, respectively.
\iffalse
{\color{red}In our numerical calculation, both $d$ and $d^*$ are composed of two
constituents (two nucleons or two $\Delta$s or two $C_8$s). Although the relation among
the angular momenta of all the constituents involved becomes much complicated, for the
process where the photon presents, the above mentioned relation remains the same as that we
discussed above after all the internal structures are integrated out. Further considering
the rule of the vector coupling at each vertex in the Feynman diagrams, it is possible to
link the helicity representation and multipole scheme.}
\fi
One can obtain the connections between the helicity amplitudes and multipole transitions from simplified physical
arguments. The $E2$ ($2^+$) transition happens when the projections of photon's spin ($J_{\gamma}=1$) and
angular momentum ($l_{\gamma}=1$ with $l_{\gamma}=L-1$ in the electric $L$-pole (EL) transition) are aligned to
that of the deuteron total spin ($m_d=+1$), and the $E4$ ($4^+$) transition appears when both the projections
of photon's spin ($J_{\gamma}=1$) and angular momenta ($l_{\gamma}=3$) are antiparallel to that of deuteron total
spin ($m_d=-1$). In addition, the $M_3$ transition is clearly associated to the $m_d=0$ state. Details for the
transition Lagrangian of $\gamma +d\to \ds$ can be referred to Ref.~\cite{Scadron:1968zz}.\\

Now, let's return to the framework of the helicity amplitudes. In the non-relativistic approximation, the matrix
elements for Fig.~{\ref{fig:Fig1}} (a,c,f) can be written as
\eq\label{eq:tranamp}
{\cal A}^{Int.}_{(a,c,f)}(m_d)&=&\sqrt{\frac{8\pi}{15}}\sum_{Int.=N,\Delta}
\frac{\big [{\mathcal M}^{\,\,\gamma N\to Int.}\big ]}{2\omega_p(2\pi)^3}\Big
[T^{Int.}\Big ]_{(a,c,f)}\Big [S^{Int.}(m_{d})\Big ]_{(a,c,f)}\nonumber \\
&&\times\int d^3p d^3k\times\Psi_d(k)\Psi^*_{d^*}\big (\frac{\vec{q}}{2}+\vec{k}+\vec{p}\big )
\Big [{\cal D}_{(a,c,f)}^{Int.}\Big ]\Big [{\cal C}_{(a,c,f)}^{Int.}\Big ]p^2Y_{20}(\Omega_p),
\en
where the superscript $"Int."$ specifies the considered intermediate state $N$ or $\Delta$, $\Psi_{d(\ds)}$
is the wave function of the deuteron ($\ds$), $\Big [T^{Int.}\Big ]_{(a,c,f)}$ represents the isospin factor
of Figs. 1(a,c,f), $\Big [S^{Int.}(m_{d})\Big ]_{(a,c,f)}$ denotes the spin factor with the projection $m_d$
of the spin of deuteron (the polarization of real photon has been chosen as $\epsilon_{\gamma}=+1$), and
$\big [{\mathcal M}^{\,\,\gamma N\to Int.} \big ]$ stands for the convention electromagnetic transition
amplitude of a three-quark nucleon in the process of $\gamma N\to N^*$, which has already been calculated
in the constituent quark model, and its explicit forms can be found in Refs.~\cite{Capstick:1992xn,
Capstick:1992uc,Close}. The energy denominators in eq.~({\ref{eq:tranamp}}) can be expressed as
\eq
{\cal D}_a^{Int.}&=&\frac{1}{[q_0+E_{d}(-\vec{q}/2)]-[E_{\Delta}^{(1)}
(3\vec{q}/4+\vec{k}-\vec{p})
+\omega_{\pi}(p)+E_{N}^{(2)}(-\vec{q}/4-\vec{k})]}\\ \nonumber
&\times&
\frac{1}{[q_0+E_{d}(-\vec{q}/2)]-[E^{Int.}(3\vec{q}/4+\vec{k})
+E_N^{(2)}(-\vec{q}/4-\vec{k})]},
\en
\iffalse
and
\fi
\eq
{\cal D}_c^{Int.}&=&\frac{1}{[q_0+E_{d}(-\vec{q}/2)]-[E^{Int.}(3\vec{q}/4+\vec{k})
+\omega_{\pi}(p)+E_{\Delta}^{(2)}(-\vec{q}/4-\vec{k}-\vec{p})]}\\
\nonumber &\times&
\frac{1}{[q_0+E_{d}(-\vec{q}/2)]-[q_0+E_N^{(1)}(-\vec{q}/4+\vec{k})
+\omega_{\pi}(\vec{p})+
E_{\Delta}^{(2)}(-\vec{q}/4-\vec{k}-\vec{p})]},
\en
and
\eq
{\cal D}_f^{Int.}&=&\frac{1}{[q_0+E_{d}(-\vec{q}/2)]-[q_0+E^{Int.}(-\vec{q}/4+\vec{k}+\vec{p})
+E_{\Delta}^{(2)}(-\vec{q}/4-\vec{k}-\vec{p})]}\\ \nonumber
&\times& \frac{1}{[q_0+E_{d}(-\vec{q}/2)]-[q_0+E_{N}^{(1)}(-\vec{q}/4+\vec{k})+\omega(\vec{p})+
E_{\Delta}^{(2)}(-\vec{q}/4-\vec{k}-\vec{p})]},
\en
respectively, and the product of the strong transition coupling of $\pi B\to B'$
in eq.~({\ref{eq:tranamp}}) for the corresponding sub-diagrams can
be denoted by
\eq
\Big [{\cal C}_{a,c}^{Int.}\Big ]&=&{\cal M}
_{N^{(2)}\to\pi\Delta^{(2)}}\times{\cal M}_{N^*\pi\to\Delta^{(1)}},~~~~~~
\Big [{\cal C}_{f}^{Int.}\Big ]={\cal M}
_{N^{(2)}\to\pi\Delta^{(2)}}\times{\cal M}_{N^{(1)}\pi\to N^*}.
\en
In  Eqs. (5-7), the superscripts "(1)" and "(2)" stand for the upper and lower $\Delta$ clusters. In
the constituent quark model, those couplings in eq. (8) have been explicitly discussed in Refs.~\cite{Close,
Riska:2000gd}. \\

Similarly, the matrix elements for the latter three subdiagrams in Figs. 1(b), 1(d), and 1(e) in  Fig.~{\ref{fig:Fig1}}
are written as
\eq
\label{eq:tranamp2}
{\cal A}^{Int.}_{(b,d,e)}(m_d)&=&\sqrt{\frac{8\pi}{15}}\sum_{I=N,\Delta}\frac{\big
[{\mathcal M}^{\gamma N\to Int.}\big ]}{2\omega(2\pi)^3}\Big
[T^{Int.}\Big ]_{(b,d,e)} \Big [S^{Int.}(m_d)\Big ]_{(b,d,e)}\nonumber \\
&&\times \int d^3p d^3k \Psi_d(k)\Psi^*_{d^*}\big
(\frac{\vec{q}}{2}+\vec{k}-\vec{p}\big )\Big [{\cal D}_{(b,d,e)}^{Int.}\Big ]
\Big [{\cal C}_{(b,d,e)}^{Int.}\Big ]p^2Y_{20}(\Omega_p),
\en
with
\eq
{\cal D}_b^{Int.}&=&\frac{1}{[q_0+E_{d}(-\vec{q}/2)]-[E^{Int.}(3\vec{q}/4+\vec{k})
+\omega_{\pi}(p)+E_{\Delta}^{(2)}(-\vec{q}/4-\vec{k}-\vec{p})]}\\ \nonumber
&\times& \frac{1}{[q_0+E_{d}(-\vec{q}/2)]-[E^{Int.}(3\vec{q}/4+\vec{k})
+E_N^{(2)}(-\vec{q}/4-\vec{k})]}, \en \eq {\cal D}_d^{Int.}
&=&\frac{1}{[q_0+E_{d}(-\vec{q}/2)]-[E_{\Delta}^{(1)}(3\vec{q}/4+\vec{k}-\vec{p})
+\omega_{\pi}(p)+E_{N}^{(2)}(-\vec{q}/4-\vec{k})]}\\ \nonumber
&\times& \frac{1}{[q_0+E_{d}(-\vec{q}/2)]-[q_0+E^{Int.}(-\vec{q}/4+\vec{k}-\vec{p})+\omega_{\pi}(\vec{p})+
E_{N}^{(2)}(-\vec{q}/4-\vec{k})]}, \en and \eq {\cal D}_e^{Int.}
&=&\frac{1}{[q_0+E_{d}(-\vec{q}/2)]-[q_0+E^{Int.}(-\vec{q}/4+\vec{k}-\vec{p})
+E_{\Delta}^{(2)}(-\vec{q}/4-\vec{k}+\vec{p})]}\\ \nonumber
&\times& \frac{1}{[q_0+E_{d}(-\vec{q}/2)]-[q_0+E^{Int.}(-\vec{q}/4+\vec{k}-\vec{p})+\omega(\vec{p})+
E_{N}^{(2)}(-\vec{q}/4-\vec{k})]}, \en and \eq \Big [{\cal C}_{b}^{Int.}\Big ]
&=&{\cal M} _{\pi N^{(2)}\to\Delta^{(2)}}\times {\cal M}_{N^*\to \pi\Delta^{(1)}},~~~~~~
\Big [{\cal C}_{d,e}^{Int.}\Big ]={\cal M} _{\pi
N^{(2)}\to\Delta^{(2)}}\times{\cal M}_{N^{(1)}\to N^*\pi},
\en
Finally, the total electromagnetic transition amplitude is summarized as
\eq
{\cal A}(m_d)=\sum_{i=a}^f\sum_{Int.=N}^{\Delta}{\cal A}_i^{Int.}(m_d).
\en
The relevant isospin and spin factors of $\big [T^{Int.}\big ]_{a,b,c,d,e,f}$ and
$\big [S^{Int.}(m_d) \big ]_{a,b,c,d,e,f}$ in eqs.~({\ref{eq:tranamp}}) and ({\ref{eq:tranamp2}}) are given in
Tables~{\ref{tab:isospin}} and {\ref{tab:spin}}. It should be mentioned that the spin factor here includes all
the necessary high-partial wave contributions due to the intermediate pion exchange in Fig. 1. For instance, in
a case with the simplest $E2$ transition and the $S$-wave $pn$ configuration of $d$, when one of the nucleon in
$d$ is excited by a photon to $N^*$ leaving another nucleon as a "spectator", the only allowed nucleon resonance
is $N^*(5/2^+)$. However, due to the pion-exchange in the time-ordered diagram, the nucleon can be excited to
$N^*(1/2^+)$ in $D$-wave to another nucleon, leaving the total spin of $d^*$ being $3$. Namely, in our numerical
calculation, the small $D$-wave component in the relative wave functions of the deuteron ($5-6\%$) and $\ds$
(about $0.5-0.6\%$ in the $\Delta\Delta$ configuration and about $0.0-0.02\%$ in the $C_8C_8$ configuration)
are explicitly taken into account (see for example Ref.~\cite{Dong:2018emq} for the $D$-wave of
$\ds$).
\renewcommand\baselinestretch{2.0}

\begin{table}[htbp]
\begin{center}
\caption{Isospin factor $\big [I^{Int.}\big ]_{(a,b,c,d,e,f)}$.}
{\begin{tabular}{||c||c|c||}\hline
%& &\\
Intermediate state &$\big [I^{Int.}\big ]_{(a,b,c)}$ &$\big [I^{Int.}\big ]_{(d,e,f)}$\\
%& & \\
\hline
%& &  \\
$Int.=N, T^{Int.}=1/2$        &$-\sqrt{2}$ &$0$\\
%& & \\
\hline
%&  & \\
$Int.=\Delta, T^{Int.}=3/2$   &$0$   &$-\frac{\sqrt{3}}{2}$\\
%& & \\
\hline
\end{tabular}
\label{tab:isospin}}
\end{center}
\end{table}

\renewcommand\baselinestretch{1.0}

\renewcommand\baselinestretch{2.0}

\begin{table}[htbp]
%\begin{center}
\caption{Spin factor $\big [S^{Int.}\big ]_{(a,b,c,d,e,f)}$ with
$B_2(m_d)=C_{1,1;2,m_d}^{3,(m_d+1)}C_{2,0;2,m_d}^{1,m_d}$ and
$B_3(m_d)=C_{1,1;3,m_d}^{3,(m_d+1)}C_{2,0;3,m_d}^{1,m_d}$,
respectively.} {\begin{tabular}{||c||c|c||}\hline
Intermediate state &$\big [S^{Int.}(m_d)\big ]_{(a,b,c)}$ &$\big [S^{Int.}(m_d)\big ]_{(d,e,f)}$\\
 \hline
$Int.=N, S^{Int.}=1/2$        &$\sqrt{\frac29}\big [B_2(m_d)
-\sqrt{2}B_3(m_d)\big ]$ &$\frac12\sqrt{\frac{10}{3}}B_2(m_d)$\\
 \hline
$Int.=\Delta, S^{int.}=3/2$   &$\frac{\sqrt{5}}{6}
\big [\sqrt{2}B_2(m_d)+\frac{8}{5}B_3(m_d)\big ]$   &$-\sqrt{\frac{28}{25}}B_3(m_d)$\\
\hline
\end{tabular}
\label{tab:spin}}
%\end{center}
\end{table}

\renewcommand\baselinestretch{1.0}

\section{Photo-absorption cross sections on deuteron in $\ds$ energy region}

\begin{figure}[htbp]
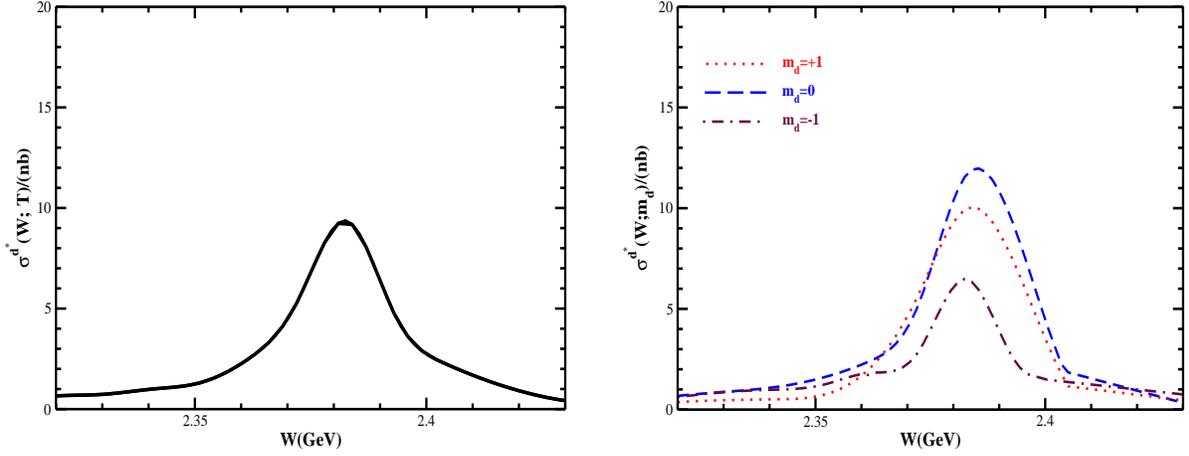

\begin{center}
\includegraphics[width=7.5cm,height=6cm]{s19sigma10.eps}
{\hskip 0.5cm}
\includegraphics[width=7.5cm,height=6cm]{s19sigma10md.eps}
\vspace{1.0cm}
\caption{Total photon-absorption cross section $\sigma^{\ds}(W;T)$ on the deuteron
target in the $\ds$ resonance region (left), and the individual contributions $\sigma^{\ds}(W;m_d)$
with $m_d=1,0,-1$ deuteron target (right).}
\label{fig:Fig2}
\end{center}
\end{figure}

It is well known that the photon-absorption cross sections on a nucleon can provide transparent information
for the nucleon resonances, such as their transition amplitudes. There have been many analyses of the real
photon-absorption cross sections in the energy region from the pion photo-production point to
$2~\rm{GeV}$~\cite{Stoler:1993yk,Stuart:1996zs,Dong:1997pv,Drechsel:2004ki,Aznauryan:2009da}. Unlike the case
of the three-quark nucleon, there are three spin-dependent photo-absorption cross sections on the polarized
deuteron with the polarization of $\sigma_{m_{d}}$ (with $m_d=\mp 1,0$), namely, they are the
absorption cross sections corresponding to the polarized photon with the helicity being antiparallel
and parallel to the transversely polarized deuteron and that to the longitudinally polarized deuteron,
respectively. Therefore, the photon-absorption cross sections ($Q^2=0$) from the contribution of a resonance
can be calculated by~\cite{Stoler:1993yk}
\eq
\label{eq:xsecres}
\sigma^{res.}(W;m_d)=\sum_{R}\frac{2M_d}{W+M_R}R(W,\Gamma)
\big |{\cal A}(W;m_d)\big |^2,
\en
where $\sum_R$ means the summation over all the resonances in the energy region, $W$ is the center-of-mass
energy, and $\Gamma$ is the total decay width of the resonance. In the low-$Q^2$ range, it is adequate to
represent the resonance shape by a simple non-relativistic Breit-Wigner form of
$R(W,\Gamma)$
\eq
R(W,\Gamma)=\frac{\Gamma}{(W-M_R)^2+\Gamma^2/4},
\en
where $M_d$, and $M_R$ stand for the masses of deuteron and $d^*(2380)$, respectively. The total
photo-absorption cross section on deuteron from the contribution of the $d^*(2380)$ resonance can be written as
\eq
\sigma^{res.}(W;T)=\frac13\Big
(\sigma^{res.}(W;m_d=+1)+\sigma^{res.}(W;m_d=0)+\sigma^{res.}(W;m_d=-1)\Big ).
\en
At the resonant point of $W=W_R$, it can be further expressed as
\eq
\sigma^{res.}(W=M_R;T)=\frac43\frac{M_d}{W_R\Gamma} \Big (|{\cal A}(M_R;m_d=+1)|^2
+|{\cal A}(M_R;m_d=0)|^2+|{\cal A}(M_R;m_d=-1)|^2\Big ).
\en
Based on the electromagnetic matrix elements obtained in Sect. III, the photo-absorption cross sections on
deuteron provided by the contribution of the $\ds$ resonance can be calculated, and corresponding results
are plotted in Fig.~{\ref{fig:Fig2}}.\\

In this figure, we find that the contribution from $\ds$ to the real photon-absorption cross sections is
in the order of ten nano-barn (nb) at the resonant point of $\ds$. The total estimated cross section is about
$\sigma^{d^*}(M_{d^*}; T)\sim 9.35~\rm{nb}$, and the $\sigma^{d^*}(M_{d^*}; m_d)$ are $9.92~\rm{nb}$,
$11.56~\rm{nb}$, and $6.49~\rm{nb}$ for $m_d=+1,0,-1$, respectively. The corresponding helicity amplitudes of
the transition are $\mid {\cal A}_1\mid=0.754\times 10^{-3}~\rm{GeV}^{-1/2}$,
$\mid {\cal A}_0\mid=0.812\times 10^{-3}~\rm{GeV}^{-1/2}$, and $\mid {\cal A}_{-1}\mid=0.608\times
10^{-3}~\rm{GeV}^{-1/2}$, respectively. Those transition amplitudes are much smaller than the helicity
amplitudes of  $\Delta(1232)$ resonance ($A_{1/2}^{\Delta}=(-141\pm 5)\times 10^{-3}~\rm{GeV}^{-1/2}$,
$A_{3/2}^{\Delta}=(-258\pm 19)\times 10^{-3}~\rm{GeV}^{-1/2}$). Moreover, the photon-absorption cross
section on deuteron from the contribution of $\ds$ in our scenario is almost $5-6$ orders of magnitude
smaller than that of about $500~\rm{\mu b}$ on nucleon from the contribution of $\Delta$ resonance (see for
example Ref.~\cite{Drechsel:2004ki}). The reasons for the remarkable suppression are the following: 1)
The lowest non-vanishing diagrams shown in Fig.~{\ref{fig:Fig1}} are in the next-leading order, since
the incoming photon excites one of the nucleon to the $\Delta$ resonance and another $\Delta$ should
emerge by exciting another nucleon via the one pion exchange, which is similar to the diagrams given
in~\cite{Kanda:2015}. 2) The wave function of $\ds$ in our suggested structural model contains two
components, $\Delta\Delta$ and $C_8C_8$, and the probability of the former is only about $1/3$ of the total.
3) Even in the $\Delta\Delta$ case, the effect that one of the $\Delta$s is bound by another $\Delta$
makes the amplitude suppressed. 4) The constituent quark model is not good enough to explain the
extracted transition amplitudes $A_{1/2,3/2}$, even for the $\Delta$ resonance, from the
experimental measurements. It can only reproduce about $2/3$ of the data. One expects that the pion
meson cloud might play an important role in understanding the data~\cite{Dong:2001js,
Kaelbermann:1983zb,Bermuth:1988ms,Lu:1997sd,Lu:1996rj,Dong:1999cz}. In addition, our obtained
${\cal A}_{\pm, 0}$ are complex numbers, unlike the real number in the tree diagram calculation for a
single nucleon excitation.\\

It should be mentioned that several experiments on the $\gamma d$ at the $\ds$ energy region have been carried out at
MAINZ~\cite{Gunther:2017ngt,Bashkanov:2018ftd}. Moreover,  an experiment on the $\gamma d\rightarrow \pi^0\pi^0d$ at
the incident energy of $0.55\sim 1.15~\rm{GeV}$ has been carried out in the Research Center for Electron Photon
Sciences (ELPH) at Tohoku University, Japan, and the new data on dibaryon were released. It is shown that a signal
of the $\ds(2380)$ state with the width of $70~\rm{MeV}$ was clearly exhibited in the mass spectrum of
$\pi^0\pi^0 d$, and the corresponding cross section at this resonant point is about $28~\rm{nb}$
~\cite{Ishikawa:2016yiq,Ishikawa:2018}. Compared with the observations, our estimated cross section  for the
total photon absorption is about $20$ times smaller than the data, since the branching ratio of
$\ds\to \pi^0+\pi^0+d$ is about $~14\%$. It should be stressed that in this calculation, we only
consider the direct coupling of a photon to one of nucleons inside the deuteron. It seems insufficient to describe
such process. This defect could be partially attributed to the following reasons. In our calculation of deuteron,
the $C_8C_8$ and $\Delta\Delta$ components are very small. Although a colorless-nucleon pair cannot be
directly converted to a colored-$C_8$ pair by a photon in the leading order approximation, a very small
hidden-color component in deuteron can change to a dominant hidden-color component in $d^*$ by the action of a
photon on the deuteron. The large change in the fraction of the hidden-color component when $d$ is converted to
$d^*$ implies that lack of this mechanism would partially affect the loss of the photo-absorption cross section.
Another reason could be that due to the existence of the external electromagnetic field, one needs the coupling of
the photon with the meson exchange current during the excitation of $d^*$  by the photon in the $pn\to \Delta\Delta$
transition~\cite{Meyer:2001js,Buchmann:1997em,Meyer:1998td,Yamauchi:1991hu,Kotlyar:1987hy}. This kind of interaction,
might provide a sizeable contribution to the $\gamma d\to d^*$ transition. Moreover, the action of a photon
on the spin-1 $\Delta\Delta$ component in deuteron can also generate the spin-3 $\Delta\Delta$ component in $d^*$.
Although the $\Delta\Delta$ component in $d$ is very small, this mechanism might also provide a measurable effect.  In
addition, due to the pion-exchange in the time-ordered diagram, the nucleon can be excited to $N^*(1/2^+)$
in $D$-wave to another nucleon, which might be another factor for the suppression of the photoabsorption cross
section compared with the data. Of course, there might also some other factors that would affect the photoabsorption
cross section, for instance other intermediate processes, the higher partial wave component in deuteron and
$\Delta$, the defect of the quark model in describing the photon associated process, and etc.. \\

Finally, in order to closely relate the theoretical results with the actual experiments, and to further obtain
the information about other intermediate mechanisms and asymmetric behaviors in the transition process of
$\gamma d \to \ds\to pn$ it is necessary to study the differential cross sections with the polarized photon
and polarized target. These will be done in our future study.\\

\vspace{1cm}

\section{Summary}
\noindent\par

In order to understand the internal structure of the $\ds$ resonance discovered by CELSIUS/WASA and WASA@COSY
Collaborations, two major structural schemes were proposed recently. One of them, based on the quark degrees of
freedom, considers that it has a compact exotic hexaquark dominated structure due to the quark exchange effect,
and the other, in terms of hadronic degrees of freedom,  believes it as a molecular-like hadronic state. These two
structures have been tested in terms of the experimental data. Up to now, both models can explain the mass, the
total width, and the partial decay widths for all the observed double pion decays of the $\ds$ resonance. However,
for a single pion decay process, although the observed upper limit of the branching ratio can be explained by
both structure models, the ways of explanation have a difference.
\iffalse
The result from a compact hexaquark dominated structure model is directly calculated and is not contradict to
the experimental data. On the other side, a combined $\alpha~[\Delta\Delta]+(1-\alpha)~[D_{12}\pi]$
mixing structure was also proposed~\cite{Gal1}, and the data can be explained as well.\fi
Therefore, we need to seek other physical quantities to distinguish these two different structures for $\ds$.  Of
course, the realistic structure of $\ds$ might be much more complicated, for instance, our compact hexaquark dominated
core as an essential ingredient is mixed with other ingredients, such as a $D_{12}\pi$ cloud~\cite{Gal1}. This picture
just looks like the commonly believed nucleon where a three-quark core is surrounded by the meson cloud. \\

The aim of this paper is to find the contribution of $d^*$ to the total photoabsorption cross section on deuteron target
by using the $d$ and $d^*$ wave functions obtained in our scenario. From a theoretical point of view, our compact
picture may not be able to provide enough contribution. This is because that in the calculated diagrams, we only
consider the $\Delta\Delta$ configuration of $d^*$, whose probability is about two times smaller than that of the
$C_8C_8$ component, since the colorless-nucleon pair cannot be converted to the colored cluster pair (hidden-color
configuration) by a photon in the lowest order approximation. Then, the estimated total photoabsorption cross section
of $\gamma d \to d^*$ at the resonant point of $d^*$ is just about ten $nb$ which is much smaller compared with the
data. Furthermore, we would mention that the quark model predictions for the helicity amplitudes of the single
$\Delta$ resonance are about $30\%$ smaller than the experimental data (see
Refs.~\cite{Dong:2001js,Kaelbermann:1983zb,Bermuth:1988ms,Lu:1997sd,Lu:1996rj,Dong:1999cz}). Even if we can make up
for the underestimation of the quark model in some way, the photoabsorption cross section of $\gamma d\rightarrow
\pi^0\pi^0 d$ from the $\Delta\Delta$ component of $\ds$ is around $1\sim 2~nb$, which is still about one order of
magnitude smaller than the currently data of about $28~\rm{nb}$ measured by ELPH and
A2 at MAINZ~\cite{Gunther:2017ngt,Bashkanov:2018ftd,Ishikawa:2016yiq,Ishikawa:2018}. The small signal might be
submerged in the contribution from the background, which comes from the other mechanisms, for instance, the conversion
of the very small hidden-color configuration in deuteron to a dominated hidden-color configuration in $d^*$, the
coupling of the photon to the meson exchange current, and so on. In particular, as pointed out in the recent
paper~\cite{Bashkanov:2019mbz}, the deformation of the $\Delta$ wave function may play an important role in the
$\gamma d \to d^*$ transition as well. Inclusion of these mechanisms would enhance the photoabsorption cross section,
so that the cross section of the $\gamma d\to d^*$ transition may become visible. It should be mentioned that although
our chiral quark model has already reasonably reproduced the deformation of wave function and the $E2/M1$
ratio for the $\Delta$ resonance~\cite{Shen:1997jd}, for roughly estimating this transition rate of $\gamma d\to d^*$,
we ignore the deformation of the $\Delta$ wave function temperedly.  It seems that this deformation effect
in the single $\Delta$ resonance is larger than that from the $D-$ wave component between the two clusters.
Therefore, those mechanisms, especially the contribution from the deformations of the single $\Delta$ clusters,
should be carefully investigated in future. In addition, we should mention that by considering the real photo-production
process, one might also obtain the information of the magnetic moment through the excitation of the $P$-shell nucleon in
nucleon pair by a photon via a $M1$ transition as well~\cite{Bashkanov:2019mbz,Bashkanov:2018}.\\

\section*{Acknowledgment}

\noindent\par
This work is supported by the National Natural Sciences Foundations of China under the  Grant Nos. 11475192,
11475181, 11521505, 11565007, and 11635009, the Sino-German CRC 110"Symmetries and the Emergence of
Structure in QCD" project by NSFC under the grant No.11621131001, the Key Research Program of Frontier
Sciences, CAS, Grant No. Y7292610K1, and the IHEP Innovation Fund under the grant No. Y4545190Y2. Authors
thank the fruitful discussions with Mikhail Bashkanov, and Yubing Dong thanks Fei Huang
for providing the wave functions of $\ds$.\\

%\newpage

\end{document}